\documentclass[a4paper, 8pt, nofootinbib, amsmath, amssymb, aps, prd, twocolumn]{revtex4-2}

\usepackage{setspace}
\newcommand{\cF}{{\cal F}}
\newcommand{\Fc}{{\cal F}}

\usepackage{CJKutf8} 
\usepackage{mathbbol} 
\usepackage{amssymb}

\usepackage[dvipsnames]{xcolor} 
\usepackage{hyperref} 
\hypersetup{
 colorlinks=true, 
 linkcolor=blue, 
 citecolor=PineGreen, 
 filecolor=magenta, 
 urlcolor=purple, 
 anchorcolor=blue,
 menucolor=Fuchsia,
 runcolor=magenta
}

\usepackage[all]{hypcap} 

\usepackage{amsmath}
\usepackage{tikz}
\usepackage{nccmath}

\newcommand{\mycomment}[1]{}
\usepackage{comment}
\usepackage{setspace}

\usepackage{graphicx}
\usepackage{dcolumn}
\usepackage{bm}
\usepackage{subcaption}

\def\ie{{\it i.e.}}

\def\tr{{\rm tr}}
\def\Tr{{\rm Tr}}
\def\Log{{\rm \,log\,}}
\newcommand{\nn}{\nonumber}

\newcommand{\be}{\begin{equation}}
\newcommand{\ee}{\end{equation}}
\newcommand{\best}{\begin{equation*}}
\newcommand{\eest}{\end{equation*}}
\newcommand{\bea}{\begin{eqnarray}}
\newcommand{\eea}{\end{eqnarray}}
\newcommand{\bear}{\begin{equation}\begin{aligned}}
\newcommand{\eear}{\end{aligned}\end{equation}}
\newcommand{\bes}{\begin{subequations}\begin{align}}
\newcommand{\ees}{\end{align}\end{subequations}}

\let\polishl=\l

\let\cwithleg=\c

\let\a=\alpha \let\b=\beta \let\g=\gamma \let\d=\delta
\let\z=\zeta     \let\l=\lambda
\let\m=\mu \let\n=\nu  \let\r=\rho
\let\s=\sigma    \let\c=\chi
  
\let\G=\Gamma \let\D=\Delta  \let\L=\Lambda

\def\cF{{\cal F}}
\def\cG{{\cal G}}

\newcommand{\na}{\nabla}

\newcommand{\ofbox}{(\Box)}
\newcommand{\ofboxstar}{(\Boxstar)}
\newcommand{\ofdelta}{\left(\D\right)}

\newcommand{\logdivsubscript}{{\rm log.\,div.}}

\newcommand{\Weyl}{C}
\newcommand{\WeylSq}{\Weyl^2}
\newcommand{\EulerInv}{E_{\rm GB}}

\newcommand{\Mstar}{M_\star}
\newcommand{\Boxstar}{\Box_\star}

\newcommand{\Rdual}{R^*\,}

\newcommand{\LUV}{\L_{\text{\tiny UV}}}
\newcommand{\Fsub}{\cF_{\rm sub}}
\newcommand{\Fsubsub}{\cF_{\rm sub-sub}}

\newcommand{\mP}{m_P}

\begin{document}
\begin{CJK*}{UTF8}{gbsn} 

\title{Cancellation of UV divergences in ghost-free infinite derivative gravity}

\author{Alexey S. Koshelev\,$^{1,2}$}
\email{askoshelev@shanghaitech.edu.cn}
\author{Oleg Melichev\,$^{1,3}$}
\email{melichev@proton.me}
\author{Les\polishl{}aw Rachwa\polishl{}\,$^{4}$}
\email{grzerach@gmail.com}

\affiliation{\vskip 2mm
$^1$
School of Physical Science and Technology, ShanghaiTech University,
393 Middle Huaxia Road, Pudong, Shanghai, 201210, China
\\$^2$
Departamento de F\'isica, Centro de Matem\'atica e Aplica\cwithleg{c}oes (CMA-UBI),
 Universidade da Beira Interior, 6200 Covilh\~a, Portugal 
\\$^3$
Yerevan Physics Institute,
2 Alikhanyan Brothers Street, 0036 Yerevan, Armenia
\\$^4$
Institute of Physics, University of Bras\'{i}lia,
Bras\'{i}lia, 70910-900, DF, Brazil}

\begin{abstract}
We consider the most general covariant gravity action up to terms that are quadratic in curvature.
These can be endowed with generic form factors, which are functions of the d’Alembert operator.
If they are chosen in a specific way as an exponent of an entire function, the theory becomes ghost-free and renormalizable at the price of non-locality.
Furthermore, according to power-counting arguments, if these functions grow sufficiently fast along the real axis, divergences may only appear at the first order in loop expansion.
Using the heat kernel technique, we compute the one-loop logarithmic divergences in the ultraviolet limit and determine the conditions under which they vanish completely, apart from the Gauss--Bonnet term and a surface term, both of which can be neglected on a four-dimensional manifold without a boundary.
We identify form factors both within the Tomboulis class and beyond it that lead to vanishing logarithmic divergences.
The general expression for the one-loop beta functions of the dimensionless couplings in quadratic gravity with asymptotically monomial form factors is given.
\end{abstract}

\maketitle

\section{\label{sec:intro}Introduction}

Infinite derivative gravity (IDG) is a modification of General Relativity (GR) motivated by the aim of ultimately resolving the two longstanding problems: non-renormalizability and the presence of ghosts.
The Einstein--Hilbert action is augmented by curvature-squared terms that involve operator functions of the covariant d’Alembertian --- also called form factors --- which serve to improve the UV behavior of the quantum theory while preserving its spectrum
\cite{Wataghin:1934ann,
Kuzmin:1989sp, 
Modesto:2011kw,
Modesto:2014lga,
Salles:2014rua, 
Tomboulis:2015gfa,  
Modesto:2016max, 
Koshelev:2016xqb, 
Koshelev:2017ebj, 
Modesto:2017uji, 
Calcagni:2022shb, 
Rachwal:2022huk, 
Rachwal:2024gjy} (for review, see \cite{Modesto:2017sdr, BasiBeneito:2022wux}).
This implies that the form factors are analytic at zero, ensuring that the correct IR limit of GR is recovered.
Their Taylor expansion must not be truncated at any finite order to avoid introducing new poles in the gravitational propagator.
This way, the spectrum of IDG contains only two dynamical metric degrees of freedom, as we consider pure gravity in the absence of matter and do not allow for any other fields. 
Non-local modifications of the electromagnetic and strong interactions are also known in the literature \cite{Krasnikov:1987yj,
Krasnikov:2024jld,
Modesto:2015foa, 
Modesto:2015lna}.

Considering the generic action at the second order in curvatures, one can write
\bear \label{action_IDG_Weyl_basis}
S \!&= \int\! d^4 x \,\sqrt{|g|}\, \Big[ 
\mP^2 \left( \L + \frac{R}{2} \right)
+ R\, \cF_R \ofbox R 
\\&
+ \Weyl^{\m\n\r\s} \cF_\Weyl \ofbox \Weyl_{\m\n\r\s} 
- \Rdual^{\m\n\r\s} \cF_E \ofbox \Rdual_{\r\s\m\n} \Big] \, ,
\eear
where $C_{\m\n\r\s}$ is the Weyl tensor and $\Rdual_{\m\n\r\s} = \tfrac{1}{2} \eta_{\m\n\a\b} R^{\a\b}{}_{\r\s}$ is the dual Riemann tensor, and $\eta_{\m\n\a\b} = \sqrt{|g|} \epsilon_{\m\n\a\b}$ is the Levi-Civita tensor.
This action depends on three yet-undefined operator functions of the d’Alembert operator $\Box = g^{\m\n} \na_\m \na_\n$.
Such an ansatz is motivated by the fact that terms cubic or of higher order in the curvature would not contribute to the tree-level two-point function around flat spacetime.
Also, any other term of the second order in curvatures with differently contracted derivatives (not forming a covariant d’Alembertian) can be rewritten as a combination of ones that are already present plus terms higher in curvatures \cite{Biswas:2011ar, Biswas:2013cha}.
A similar-looking action commonly appears in the context of asymptotically safe gravity \cite{Draper:2020knh,
Knorr:2022dsx,
Knorr:2022lzn,
Knorr:2023usb}, as a curvature expansion of the average effective action; as well as in one-loop calculations \cite{Teixeira:2020kew,
Ohta:2020bsc}.
Here, however, we treat \eqref{action_IDG_Weyl_basis} as the starting classical action of our theory.

The last term in \eqref{action_IDG_Weyl_basis} represents a generalization of the Euler--Gauss--Bonnet invariant:
\bea \label{Gauss-Bonnet-generalized-nonlocal}
&& - \Rdual^{\m\n\r\s} \cF_E \ofbox \Rdual_{\r\s\m\n} 
\\&&
= R \,\cF_E \ofbox R
- 4 R^{\m\n} \cF_E \ofbox R_{\m\n}
+ R^{\m\n\r\s} \cF_E \ofbox R_{\m\n\r\s} 
\, .
\nn
\eea
This non-local term with a non-constant form factor is no longer topological. 
It turns out that it does not contribute to the tree-level two-point functions around flat spacetime (see Appendix~\ref{sec:app:Ghost-free-action}), but it will contribute to the scattering amplitudes and the beta functions.
This is also a manifestation of the fact that this term can be re-expressed in a form that is third order (and higher) in curvatures.
If we were more pedantic about treating the action as curvature expansion, we could have neglected it alongside other higher curvature terms.
However, as we will find out later, the inclusion of this term is helpful for the cancellation of divergences generated by other terms.

Throughout the paper, we restrict attention to the case of a vanishing cosmological constant, $\L=0$; its appearance earlier was merely for completeness.
The demand that the spectrum around Minkowski spacetime contain only one particle --- the massless spin-2 graviton ---
can be met by choosing form factors $\Fc_R\ofbox$ and $\Fc_C\ofbox$ to be proportional to each other and satisfy the relation (for details, see Appendix~\ref{sec:app:Ghost-free-action}):
\be \label{ghost-free-no-sclarals-condition}
x \equiv \Fc_\Weyl\ofbox / \Fc_R\ofbox = - 3 \, ,
\ee
while having a particular form:
\be \label{ff_as_exp}
\Fc \ofbox \equiv \Fc_R \ofbox 
= \mP^{2} \frac{1- e^{2 \omega \ofboxstar}}{12 \Box} \, ;
\ee
where the operator function $\omega \ofboxstar$ is an entire function of $\Boxstar = \Box/\Mstar^2$, with $\Mstar$ serving as the energy scale around which non-local effects become important.

After fixing a particular minimal (De Donder) gauge, the graviton propagator reduces to the standard GR graviton propagator acted on by the operator $e^{ - 2 \omega \ofboxstar}$. 
This operator has no zero modes and thus does not give rise to new poles on the complex plane.
Hence, the perturbative spectrum of quantum fluctuations around flat spacetime is exactly the same as in GR.
The third form factor $\Fc_E \ofbox$ has no impact on the spectrum and therefore is not restricted by the consideration of the physical degrees of freedom.

From the quantum field theory point of view, one can show that form factors with a power-law asymptotics along the positive real axis are favored. 
Such asymptotic behavior allows one to implement standard power-counting reasoning
so that these form factors can give rise to super-renormalizable theories  
\cite{Kuzmin:1989sp, Tomboulis:1997gg, Modesto:2011kw}.
In the UV limit, these theories resemble higher-derivative models with a finite number of derivatives.
However, in theories with form factors enjoying faster growing UV asymptotics (for example, exponential), the problem of renormalizability is still unclear since power-counting is not well defined in this case \cite{Salles:2014rua}. 
Moreover, it was shown for a scalar field theory that infinite derivative form factors growing faster than a power-law at infinity lead to a strong coupling regime \cite{Koshelev:2021orf}.
In this regard, we will consider power-law high-momenta asymptotics of the form factor $\Fc\ofbox$ defined in \eqref{ff_as_exp}
which we fix as
\be \label{F_asympt_q}
\cF \ofbox \rightarrow  A\, \Boxstar^q \, ,\quad \text{for}\; k \gg \Mstar \, ;
\ee
for some positive real $q$. 
$A$ is a constant that will drop out of the final result.
Even though the form factor $\Fc_E(\Box)$ does not affect the propagator, it does contribute to vertices, and therefore, by power-counting arguments, it should not grow faster than the other form factors. 
In principle, a slower growth is also allowed.
However, for the sake of simplicity, we choose it to be proportional to the other form factors as well and define
\be \label{y-definition}
y \equiv - \Fc_E \ofbox / \Fc_R \ofbox \, .
\ee
Therefore, our analysis utilizes  two independent relative parameters $x$, $y$, and one form factor $\cF \ofbox$ (mainly its asymptotics) defined in (\ref{ghost-free-no-sclarals-condition}), (\ref{y-definition}), (\ref{ff_as_exp}), and (\ref{F_asympt_q}) respectively. Even though $x$ is fixed by the condition of the absence of new degrees of freedom to be $-3$ (in 4 dimensions) we will keep it arbitrary throughout most of the computations.

The superficial degree of divergence a Feynman graph $\cG$ is then calculated to be
\be
\d (\cG) = 4+2q(1-L) \, ,
\ee
where $L$ is the number of loops.
This leads immediately to the conclusion that, for positive $q$, the superficial degree of divergence becomes negative at higher loops.
For $q > 2$, only the first loop can be divergent \cite{Modesto:2011kw, Modesto:2014lga}.
Note that this only counts the powers of momenta and does not exclude the possibility of divergent subgraphs.
Their presence cannot be ruled out and would require the computation of higher-loop corrections, which is beyond the scope of the present paper.
While this has been known for a long time, several authors have expressed doubts on whether constructing a finite or even a renormalizable unitary theory is feasible after all (cf. \cite{Asorey:2018wot, Vale:2025nef}).
In this paper, we make progress toward a positive answer to this complicated, long-standing problem.

To this end, we study the quantum corrections generated by the action \eqref{action_IDG_Weyl_basis}.
The construction of entire operator functions $\omega \ofboxstar$ entering \eqref{ff_as_exp}, satisfying the power-law UV asymptotics along the real line described above, was first achieved by Kuzmin \cite{Kuzmin:1989sp} and independently by Tomboulis \cite{Tomboulis:1997gg, Tomboulis:2015gfa, Tomboulis:2015esa}.
We will argue that, for Tomboulis-like form factors, the beta functions are determined solely by the UV asymptotics of $\cF$.
This greatly simplifies the analysis, effectively reducing it to a study of power-law-like form factors. 
We will further extend this class of operator functions, while still allowing the computation of beta functions to rely only on the asymptotic behavior of the form factors.

Within the background field method, one can fix the covariant structure of the logarithmically divergent part of the effective action as
\bear \label{log_div_structure}
&\G_\logdivsubscript\!
= \frac{1}{32 \pi^2} \log \left( \frac{\LUV^2}{\m^2} \right) 
\times \\& \times
\int\! d^4 x \;\sqrt{|g|}\, 
\left[ 
b_R R^2 
+ b_\Weyl \WeylSq 
+ b_E \EulerInv 
+ b_D \Box R
\right] \, ,
\eear
where $\LUV$ is the UV cutoff scale, $\m$ is the running scale, $\WeylSq = \Weyl^{\m\n\r\s} \Weyl_{\m\n\r\s}$, and $\EulerInv = R^{\m\n\r\s} R_{\m\n\r\s} - 4 R^{\m\n} R_{\m\n} + R^2$ is the topological Euler--Gauss--Bonnet invariant in 4 dimensions. 
Above, all the curvature tensors and covariant derivatives are taken with respect to the background metric. 
The last contribution, $\Box R$, will be ignored as a total derivative; hence, we will not compute the coefficient $b_D$.
The existence and definitions of quadratic and quartic divergences are scheme-dependent, and we will not compute them either.

In the rest of the paper, we compute the one-loop logarithmic divergences generated by the action \eqref{action_IDG_Weyl_basis} to determine the coefficients $b_\Weyl$, $b_R$, and $b_E$ that enter \eqref{log_div_structure}.
After discussing the general structure of the kinetic operator, we will compute the result by choosing a power-law form factor in section \ref{subsec:div_power_law} with some technical details collected in Appendix~\ref{sec:app:intermediate}.
We then argue that the same result applies to a certain class of form factors that are exponential of an entire function with a power-law UV asymptotics.
These considerations are presented in section \ref{subsec:general_ff}, and an explicit proof is detailed in Appendix~\ref{sec:app:integrals}.
We derive the conditions under which ghost-free gravity exhibits vanishing logarithmic divergences in section \ref{sec:discuss}.
We summarize our findings at the end of that section.

\section{\label{sec:div_com}Computation of divergences}

We work in the Euclidean signature, in which the actions \eqref{action_IDG_Weyl_basis} and \eqref{log_div_structure} acquire an overall minus sign, except for the sign in front of the last term in \eqref{action_IDG_Weyl_basis}, which is preserved.
The background field method yields the following expression for the one-loop effective action
\cite{Feynman:1963ax,
DeWitt:1967yk, Faddeev:1967fc,
Buchbinder:1992rb}:
\be \label{1-loop_EA_with_ghosts}
\G^{\rm 1-loop} = \frac{1}{2} \Tr\Log \mathbb{H} - \Tr\Log \Delta_{{\rm gh}} - \frac{1}{2} \Tr\Log \mathbb{C} \, .
\ee
Here we select the quantum variable --- the one to be integrated over in the path integral --- as the linear metric fluctuation $h_{\m\n} = {\sqrt{2}}/{\mP} \left( g_{\m\n}-\bar{g}_{\m\n} \right)$, where $\bar{g}_{\m\n}$ is the background metric field.
The first term constitutes the Hessian of the gauge-fixed action, while the second and third terms account for the Faddeev--Popov ghosts and the third ghost operator, respectively.
Their forms will be determined later.
It is imperative that the operators $\mathbb{H}$, $\D_{\rm gh}$, and $\mathbb{C}$ are self-adjoint.\footnote{
There is a subtlety related to the definition of the kinetic operator that requires a configuration space metric. 
One can choose it to be ultra-local, and any related arbitrariness drops out of the final result.  
The most convenient choice for the inverse metric will be $\mathbb{K}$ in  \eqref{factorization_of_K_from_Hessian}.
}
The diffeomorphism invariance-breaking term can be selected as
\be \label{Sgf}
S_{\rm gf} = \frac{1}{2 \a} \int\! d^4 x \sqrt{g}\, \chi_\m \mathbb{C}^{\m\n} \chi_\n \, ,
\ee
where 
\be \label{gaugefix_cond}
\chi_\m = \na^\l h_{\l\m} - \b \na_\m h \, 
\ee
is the gauge-fixing condition and
\be \label{3rd_ghost_op}
\mathbb{C}^{\m\n} = - g^{\m\n} \Box \cF\ofbox + (1-\g) \na^\m \cF\ofbox \na^\n \, 
\ee
is a self-adjoint differential operator that is chosen in such a way that the gauge-fixing action \eqref{Sgf} is of the same order in derivatives as the original action \eqref{action_IDG_Weyl_basis}, hence having the same high-momentum behavior.
The operator \eqref{3rd_ghost_op} enters the last term in \eqref{1-loop_EA_with_ghosts} as the third ghost operator.

An important technical simplification can be achieved by restricting to covariantly constant backgrounds, defined by the condition $\na_\a R_{\m\n\r\s} = 0$.
One can see from \eqref{log_div_structure} that this will be sufficient in order to compute the coefficients $b_R$, $b_\Weyl$, $b_E$.

With the UV asymptotics \eqref{F_asympt_q}, the second variation of the action \eqref{action_IDG_Weyl_basis} together with the gauge-fixing term \eqref{Sgf}  
becomes a differential operator of the order $2q+4$ (in the number of derivatives) in the UV limit.
The full expression for the Hessian has quite a complicated structure that we will not display here.
It is crucial to notice, however, that the condition of proportionality of the form factors in action \eqref{action_IDG_Weyl_basis}, together with an appropriate choice of the gauge-fixing parameters $\a$, $\b$, and $\g$, allows us to recast the Hessian into the minimal form.
We call a differential operator minimal when its principal (= highest order in derivatives) part contains derivatives that are all contracted with each other, and therefore produce covariant d'Alembert operators.
Indeed, by choosing
\best 
\a = \frac{1}{2x} 
\, , \quad
\b = \frac{x-6}{4x-6} 
\, , \quad 
\g = \frac{2}{3} - \frac{1}{x} \, ,
\eest   
we obtain
\be \label{factorization_of_K_from_Hessian}
\mathbb{H} = \mathbb{K} \left[ \Box^2 \cF\ofbox \mathbb{1} + \mathbb{M} (R, \Box) \right] \, ,
\ee
where $\mathbb{K}$ is a function of the metric that can be cast in a form that diagonalizes the first term inside the brackets.
The operator $\mathbb{M}$ is subleading with respect to the first term for high-frequency perturbations (where the number of derivatives is important), which allows us to expand the $\Tr\log$ in a Taylor series:
\bear \label{Tr_log_H_as_Taylor_series}
&\Tr\log \mathbb{H} = 2\,\Tr\log \Box + \Tr\log \cF\ofbox 
\\&
+ \Tr \Big[ \mathbb{M}\,\Box^{-2} \cF^{-1}\ofbox - \frac{1}{2} \Big( \mathbb{M}\,\Box^{-2} \cF^{-1}\ofbox \Big)^2 + \dots \Big] \, . 
\eear
The terms in the ellipsis are at least cubic in the curvature and therefore do not contribute to the divergent part, in accordance with power-counting arguments and the form of the counterterms in \eqref{log_div_structure}.
At this point, we need to compute the functional traces on the right-hand side of \eqref{Tr_log_H_as_Taylor_series}, which is done by specifying the form factor $\Fc(\Box)$.

\subsection{\label{subsec:div_power_law}Monomial Form Factors}

Consider first the power-law case:
\be \label{ff-power-law}
\cF \ofbox = \Boxstar^q \, ,
\ee
where $\Boxstar = \Box/\Mstar^2$ as before and $q$ is a positive integer larger than one.
We compute the second variation of the action \eqref{action_IDG_Weyl_basis} to get the Hessian.
Then, integrating by parts, one may write:
\bear \label{var_RR}
& \d^2 \Big\{ \int\! d^4 x\,\sqrt{g}\,R\, \Box^q R \Big\}
\\&= 2 \int\! d^4 x\,\sqrt{g}\, 
\Big\{ 
\Big[ \na^\a  \d \left( \na_\a  R \right) \Big] \,\Box^{q-2}\,
\na^\b \d \left( \na_\b R \right)
\Big\} \, ,
\eear
and analogously for the other two terms. 
Contributions containing derivatives of any of the curvature tensors (Riemann tensor, Ricci tensor, or Ricci scalar) vanish on covariantly constant backgrounds.
The operator $\Box^{q-2}$ in \eqref{var_RR} is local for $q\geqslant 2$, and one can notice that \eqref{var_RR} naturally yields a self-adjoint operator.

In order to compute the functional traces, we simplify tensorial structures by symmetrizing covariant derivatives, and use the technique of universal functional traces \cite{Barvinsky:1985an, Groh:2011dw}, which can be formulated as the following compact expression:
\begin{align}
\label{universal_functional_trace_final}
&\Tr \left[\na_{(\m_1} \dots \na_{\m_N)} f(\D) \right] 
\\&
= \frac{1}{\left(4\pi\right)^{d/2}} \sum_{n\geqslant 0} Q_{-n+\frac{d}{2}+\lfloor N/2 \rfloor} [f] \cdot
\tr \!\int\! d^d x\, \sqrt{g}\, K_{(\m_1 \dots \m_N)}^{\;(n)} \, . \nonumber
\end{align}
Here, the traces on the left-hand side are functional traces, while those on the right are traces over Lorentz indices. 
Functions $f$ depend on $\D = - \Box$ (which is a positive-definite operator in Euclidean signature), $\lfloor N/2 \rfloor$ denotes the floor function applied to the number of uncontracted derivatives on the left-hand side, $K^{(n)} (x)$ are certain local invariants constructed of the curvature tensors and their derivatives listed in \eqref{K_heat_kernel_invariants}, $d$ is the number of spacetime dimensions, and symmetrization over all indices is understood.
The $Q$-functionals are momentum integrals; their definition can be found in the Appendix \ref{sec:app:intermediate}, alongside some intermediate results.
For the power-law case \eqref{ff-power-law}, one will get the traces of such type with $f(\D) = 1 / \D^l$ with $l>0$, and the logarithmic divergences can be extracted by collecting the coefficients in front of the $1/\D$ terms inside the momentum integrands.
The final result for the coefficients entering \eqref{log_div_structure} reads
\begingroup
\allowdisplaybreaks
\bea \label{log_div_result}
b_\Weyl &&= \frac{1}{9720 x^2} \Big\{ 
-540 x^3 +18 x^2 \left(5 y^2+90 y+1071\right)
\nn\\* &&
+ 540 x y (y-48) + 15390 y^2
+ q \left[6 x^2 \left(20 y^2+345 y
\right.\right.\nn\\* &&\left.\left.
-6939\right)
+270 x y (4 y-1)+28080 y^2\right]
\nn\\* &&
+ q^2 \left[5 x^2 \left(8 y^2+198 y-5103\right)
\right.\\* &&\left.
+30 x y (38 y+423)+25740 y^2\right]
\Big\} \, ,
\nn\\  
b_R &&= \frac{1}{2916 x^2} \Big\{
9 x^2 (y+9)^2+1620 x y+405 y^2
\nn\\* &&
+ q \left[-6 x^2 \left(2 y^2+39 y+27\right)+1890 x y-540 y^2\right]
\nn\\* &&
+ q^2 \left[4 x^2 y (5 y+36)+30 x y (y+63)+1530 y^2\right]
\Big\} \, ,
\nn\\  
b_E &&= - \frac{1}{4860 x^2} \Big\{
9 x^2 \!\left(5 y^2+90 y+2352\right)+4050 x y+2025 y^2
\nn\\* && 
+ q \left[3 x^2 \!\left(20 y^2+120 y+5427\right)+27000 x y+2700 y^2\right]
\nn\\* &&
+ q^2 \left[20 x^2 y (y+18)+300 x y (y+36)+7200 y^2\right] 
\Big\} \, , \nn
\eea
\endgroup
The corresponding beta functions are obtained by taking the logarithmic derivative of \eqref{log_div_structure} with respect to the running scale $\m$.
Note that \eqref{var_RR} is invalid for $q<2$ on covariantly constant backgrounds, and hence also the results \eqref{log_div_result}.
The beta functions corresponding to  $q=1$ have been computed earlier in \cite{Rachwal:2021bgb}.
Note that the theory considered so far is described by the action \eqref{action_IDG_Weyl_basis} with the power-law form factor \eqref{ff-power-law} with integer $q$ greater than one.
Although there is some theoretical interest in this model \cite{Vale:2025nef}, it suffers from the ghost issue.

\subsection{\label{subsec:general_ff}More General Analytic Form Factor}

A general form factor can be represented as an infinite series:
\be \label{F_as_series}
\cF \ofbox = f_0 + f_1 \Boxstar + \sum_{n=2}^\infty f_n \Boxstar^n \, .
\ee
We want to compute the second variation of the action \eqref{action_IDG_Weyl_basis}.
Notice that the Hessian is a linear function of the action, but the effective action is not.
Therefore, if we want to use the representation \eqref{F_as_series}, it is desirable to perform the subsequent resummation at the level of the Hessian. Namely, we write
\be \label{Hess-as-series}
H^{\rm IDG} = H_{\rm EH} + f_0 H_0 + f_1 H_1 + \sum_{n=2}^\infty f_n H_n \left( \Boxstar \right) \, ,
\ee
where $H_{\rm EH}$ is the second variation of the Einstein--Hilbert action and $H_n$ are the Hessians of the corresponding Taylor expansion terms, each one of them being a power-law form factor $\cF \ofbox = \Boxstar^n$.
In the previous section, where such monomial form factors were considered, we computed the Hessian corresponding to $H_n$ for $n \geqslant 2$. 
After that, all tensorial manipulations leading to \eqref{log_div_result} can be repeated, keeping the form factor arbitrary, assuming only its analyticity.
The difficulty appears at the very last step, when the only tensorial structures left are the ones of the types \eqref{log_div_structure}, and one has to perform the integration over momenta.
For generic form factors \eqref{ff_as_exp}, one obtains nine integrals, each depending on a particular combination of the form factor and its derivatives \eqref{integrals_sum}, \eqref{integrals_list}.
When the form factor is not specified, it may be hard or impossible to compute some of those integrals.

However, this complication can be partially overcome by imposing certain conditions on the form factor asymptotics.
Remember that we consider form factors with power-law asymptotics along the positive real axis.
On the other hand, from the representation \eqref{ff_as_exp}, the form factor for large momenta can be written as $\cF \sim - \exp \left(2\omega(z) \right)/(12z)$ where $\omega(z)$
, which we also call the form factor logarithm,
must be an entire function to satisfy the unitarity requirement.
Here we use $z$ for eigenvalues of the positive definite operator $\D = - \Box$ (in Euclidean signature).
This means that $\omega(z)$ behaves logarithmically along the positive real line. 
For an entire function to behave logarithmically along some direction, it should be of an exponential type at least, \ie, its maximal growth rate on the complex plane should be given by $\exp(sz^\rho)$ for some $\rho\geqslant 1$. 
This in turn implies that the form factor $\cF(z)$ itself should be an entire function of an infinite order, \ie, its maximal growth is at least $\exp(e^{z})$. 
This is in particular true for form factors suggested by Tomboulis where an explicit form of the form factor logarithm is $\omega(z)=\Gamma(0,z^{q+1})+\gamma+\log(z^{q+1})$ where $\Gamma(0,z^{q+1})$ is an incomplete Gamma-function (equivalently, it can be written via exponential integral of the first kind as $\text{Ei}_1(z^{q+1})$), and $\gamma$ is the Euler--Mascheroni constant. The resulting asymptotics of the form factor is $\cF\sim A z^q$. 
A remarkable observation at this point is that at infinity $\omega(z)$ has an expansion $\gamma+\log(z)+O(e^{-z}/z)$, which means that all corrections to the power-law asymptotics of $\exp \left( 2 \omega (z) \right)$ are double-exponentially suppressed.
As discussed in the Appendix \ref{sec:app:integrals},
using only the leading asymptotic term of the form factor expansion at infinity in computations of the beta functions is justified to get the complete result, for all form factors that do not have subleading asymptotics higher than $1/z$.
In other words, for Tomboulis-like form factors, only the leading UV asymptotics contributes to the divergences.

Hence, we conclude that the result \eqref{log_div_result} initially derived in the previous subsection for a monomial form factor is also applicable for the asymptotically monomial Tomboulis form factor.
Applying the ghost-free condition $x=-3$ defined by the Eq. \eqref{ghost-free-no-sclarals-condition} to \eqref{log_div_result}, we obtain
\begingroup
\allowdisplaybreaks
\bea \label{log_div_result_no_ghosts}
b_\Weyl &&= q^2 \left(\frac{7 y^2}{27}-\frac{y}{3}-\frac{21}{8}\right)
+ q \left(\frac{8 y^2}{27}+\frac{2 y}{9}-\frac{257}{60}\right)
\nn\\*&&
+\frac{y^2}{6}+\frac{19 y}{18}+\frac{43}{20} 
\, , \quad\quad
\nn\\
b_R &&= \frac{1}{324} \Big[2 q^2 y (10 y-27)-2 q \left(4 y^2+48 y+9\right)
\\*&&
+6 y^2-42 y+81\Big]
\, , \quad\quad
\nn\\
b_E &&= \frac{1}{540} \Big[-40 q^2 y (2 y-9)+q \left(-40 y^2+960 y-1809\right)
\nn\\*&&
-6 \left(5 y^2-10 y+392\right)\Big]
\, ,\nn
\eea
\endgroup
This is our main result.

We can extend it to another class of form factors.
In doing so, we can use the Taylor series expansion. We note that the Tomboulis function $\omega(z)$ for $q=0$ can be represented as $-\sum\limits_1^\infty (-z)^n/(n!n)$. 
The radius of convergence of a series for an entire function is infinite; therefore, any extra factors in the denominators of the series terms will retain the function to be entire. 
Two explicit series can be constructed; one of them is given by:
\be \label{omega_example_Alexey_1}
\omega(z) 
= -\sum\limits_1^\infty \frac{(-z)^n}{n!^2 n} 
= z{}_2H_3(1,1;2,2,2;-z) \, ,
\ee
which can be represented at infinity as $2\gamma+\log(z)+O((\cos(2\sqrt{z}-\sin(2\sqrt{z})/z^{3/4})$. Here ${}_2H_3(z)$ is the generalized hypergeometric function. Another is given by 
\be \label{omega_example_Alexey_2}
\omega(z) 
= -\sum\limits_1^\infty \frac{(-z)^n}{(2n)!n}
= 2\gamma+\log(z)-2\text{Ci}(\sqrt{z}) \, ,
\ee
which can be represented at infinity as $2\gamma+\log(z)+O(\sin(\sqrt{z})/\sqrt{z})$. Here $\text{Ci}(z)$ is the integral cosine function. 
Both clearly yield logarithmic asymptotics, although the subleading terms are suppressed less than exponentially. 
However, we are interested in the exponent of $\omega$, which means that
the subleading contributions to the power-law asymptotics of $\exp(2\omega(z))$ remain exponentially suppressed.
Replacing in any of the above series $z\to z^{q+1}$ makes $\Fc(z)$ asymptote to $z^q$.

As can be seen from \eqref{ff_as_exp}, the form factor will always have a subleading UV asymptotics of $1/z$ in addition to the leading one of $z^q$.
It remains an open question whether one can construct a form factor obeying the unitarity condition \eqref{ff_as_exp} with a power-law asymptotic behavior, but its subleading terms at infinity are suppressed more strongly than $1/z$.
Computations of divergences and corresponding beta functions for form factors containing other subleading asymptotics, such as $z^{q-1}$, if relevant, can be obtained by applying the reasoning of Appendix \ref{sec:app:integrals}.
This would require an additional study, but some relevant results can be found in \cite{Asorey:1996hz,Modesto:2017hzl}.

\section{\label{sec:discuss}Cancellation of Divergences}

We have computed the logarithmic divergences generated by the action \eqref{action_IDG_Weyl_basis} at the one-loop order using the momenta cutoff regularization scheme.
If one were to use the dimensional regularization instead, the result would be the same, while quadratic and quartic divergences would be absent.
The result is also independent of the gauge choice, as can be seen by a straightforward repetition of the arguments used in a similar computation done in Quadratic Gravity \cite{Buchbinder:2021wzv}.
Furthermore, if the UV asymptotics of the form factor is sufficiently high, loops of the second and higher orders all become finite by power-counting.

The computation presented in this paper is the first of this type, apart from the one performed in \cite{Kuzmin:1989sp}, which remained largely unnoticed for a long time.
We believe that it contained several important conceptual ideas novel for its time, but the final result appears quantitatively erroneous.
It is difficult to identify possible mistakes, as both the paper \cite{Kuzmin:1989sp} and the author’s PhD thesis lack intermediate formulae.
Our paper presents an alternative computation using the heat kernel technique, which is perhaps clearer.

Let us see if it is possible to achieve a cancellation of the divergences computed in the previous section.
Recall that to have a ghost-free theory, we must satisfy the condition $x=-3$, which prevents generating additional poles apart from the massless graviton.
Also, $q > 2$ is necessary for higher loops to be finite.
For the finiteness of the first loop, we would need to solve the system of three independent equations $b_\Weyl = b_R = b_E = 0$.
Unsurprisingly, they admit no solution for $q$ and $y$ with a fixed $x$, because we have three equations and only two independent variables: $q$ and $y$.
However, if we allow for the remaining divergence to be of the form of the Euler--Gauss--Bonnet invariant $\EulerInv$, we can relax the corresponding system only to solve the two equations $b_\Weyl = b_R = 0$. 
Fortunately, coupled with the condition $q \geqslant 2$, which was demanded to derive our results, they do admit a unique real solution
\be 
q \approx 6.455902 \, ,~~ y \approx 3.708353 \, .
\label{sols}
\ee
On this solution, the $\WeylSq$ and $R^2$ terms are absent from the Eq. \eqref{log_div_structure}, while the remaining divergence of the Gauss--Bonnet type with $b_E \approx 27.777086$ is topological and therefore can be neglected on a four-dimensional manifold without a boundary.

When the boundary term is of importance, one can treat it simply by augmenting the action \eqref{action_IDG_Weyl_basis} with an additional Euler--Gauss--Bonnet term without any form factors, but with a new (inverse) coupling in front that we call $\r$.
This will not affect the divergences provided $q\geqslant 2$, but instead allow for the reabsorption of the corresponding counter-term.
In other words, defining a new form factor $\tilde{\cF}_E \ofbox = \cF_E \ofbox - 1/\r$, and defining the corresponding beta function from \eqref{log_div_result_no_ghosts} as
\be
\b_\r = - \frac{\r^2}{16 \pi^2} b_E 
\approx - \frac{27.78 \,\r^2}{16 \pi^2} 
\approx -0.17590 \r^2\, ,
\ee
we are also able to renormalize the topological term.
The negative sign tells that the coupling $\r$ tends to zero in the UV limit, in analogy with asymptotic freedom of quadratic gravity.
In principle, the same procedure can be done for another surface term $\Box R$, which we, however, have neglected.

Another interesting question is whether or not one can preserve the value of the parameter $x$ at the quantum level.
For this, we have to demand $q>2$ for the finiteness of higher loops and $b_\Weyl / b_R = x = - 3$.
In particular, for $q=3$ we obtain two solutions
$y \approx -1.88694$ and $y \approx 3.69376$.

In passing, we also note that from the expression \eqref{log_div_result}, and without the assumption $x=-3$, one can get a completely UV-finite theory (without the divergence proportional to the topological term $R^*R^*$), though non-unitary, if one chooses the following values of the relative couplings $x$, and $y$, and the exponent $q$ as ones from the following three sets of real solutions: $\{q \approx 2.09879,\,x \approx 11.7883,\,y \approx -11.485\},$ $\{q \approx 2.9706,\,x \approx -6.95091,\,y \approx 7.28367\}$ or $\{q \approx 12.0035,\,x \approx -9181.64,\,y \approx -6.6463\}$.

We summarize the findings and limitations of this work as follows.
We consider the theory described by the Lagrangian containing all terms that contribute to the propagator near flat space, but not other higher-curvature terms such as $R^4$ or the Goroff--Sagnotti term.
For the considered class of form factors, the result \eqref{log_div_result} is universal and exact in the UV limit, provided higher loops are finite without divergent subgraphs. This is plausible for the solution in (\ref{sols}), which removes one-loop divergences for boundary-less four-dimensional space-times.
One should keep in mind that inclusion of higher-curvature terms into the action would alter the results for the beta functions, potentially allowing for a larger space of UV-finite theories.
Second, we assumed that the form factors are proportional to each other and have a power-law UV asymptotic, or more precisely, $A \Box^q + O(1/\Box)$.
This is true, in particular, for all form factors discussed in the review \cite{Modesto:2017sdr}.
Under these conditions, the demands of the theory to be ghost-free around Minkowski and for the logarithmic divergences to vanish fix $q$, as well as the relative ratios $x$ and $y$ between the form factors, uniquely.

An important and interesting question for the studied class of theories is the UV behavior of amplitudes, which is awaiting its resolution in forthcoming papers. It is not excluded that a desired growth of an amplitude justified by unitarity and causality bounds may be in tension with the solution in (\ref{sols}), because the value of $q$ gives naively a higher than permitted UV amplitude asymptotics. 
This will prompt for finding other configurations that make the 1-loop corrections finite. Additionally, an extra adjustment of our computation method which is not valid for $q<2$ may be required. 
It is not excluded a priori that the theory will not be finite in this case.

\section*{\label{sec:Acknowledgments}Acknowledgments}

AK would like to thank Anna Tokareva for multiple discussions during the work on this manuscript.
OM is indebted to Kevin Falls for multiple insightful conversations and for hosting him for a short period at the University of the Republic in Montevideo during the preparation of this paper, as well as to Roberto Percacci and Abhishek Naskar for helpful suggestions.
AK and OM are thankful to Sian Yang for his initial collaboration on the project.
The work of OM has been supported by the Armenian HESC grant number 24RL--1CO47.

\begin{appendix}

\section{\label{sec:app:Ghost-free-action}Ghost-free action}

In this section, we derive the conditions for the absence of ghosts.
The starting action of Infinite Derivative Gravity can be written as
\bear
\label{action_IDG_Riemann_no_CC}
S \!&= \int\! d^4 x \,\sqrt{|g|}\, \Big[ 
\frac{\mP^2}{2} R  
+ R \cF_1 \ofbox R 
\\&
+ R^{\m\n} \cF_2 \ofbox R_{\m\n} 
+ R^{\m\n\r\s} \cF_3 \ofbox R_{\m\n\r\s} 
\Big] \, ,
\eear
which is equivalent to \eqref{action_IDG_Weyl_basis} for a vanishing cosmological constant, with the following relation between form factors
\bear
&\Fc_1 \ofbox = \frac13 \Fc_\Weyl \ofbox + \Fc_E \ofbox + \Fc_R \ofbox \, ,
\\&
\Fc_2 \ofbox = - 2 \Fc_\Weyl \ofbox - 4 \Fc_E \ofbox \, ,
\\&
\Fc_3 \ofbox = \Fc_\Weyl \ofbox + \Fc_E \ofbox \, .
\eear
The full nonlinear equations of motion for this action have been derived in \cite{Biswas:2013cha}.
They have a very complicated structure as a double infinite series.
However, as we are interested in the spectrum around the Minkowski background, we will only need the part that is linear in curvatures.
It reads
\bea \nn
&&E^\m_\n = \mP^{2}G^\m_\n
- 4\bigg[ \left( \na^\m\nabla_\n-\d^\m_{\n} \Box \right) \Fc_1 \ofbox R 
\\&&
+\left(-\frac12 \d^\m_\s\d_\n^\rho \Box+\d^\m_\s\na^\rho\na_\n-\frac12\d^\m_\n\na_\s\na^\rho\right) \Fc_2 \ofbox R^\s{}_\r
\nn\\&&
+ 2\, \na_\a \na_\b \Fc_3 \ofbox R_\n{}^{\a\b\m} \bigg]+O(R^2) = 0 \, .
\nn
\eea
Continuing for the Minkowski background, which allows us to move derivatives freely, and using a property
\best
\na^{\a}\na^{\b} \Weyl_{\a\m\n\b}  =-\Box S_{\m\n}+\frac{1}{6}\na_{\m}\nabla_{\n}R+O(R^2)
\, ,
\eest
where $S_{\m\n}=\frac12(R_{\m\n}-\frac16Rg_{\m\n})$ is the Schouten tensor, we can recast equations of motion as follows
\bea \label{IDG_EOM_linearized_Riemann_basis}
&&E^\m_\n = \left[\mP^{2}+2\Box\left(\Fc_2\ofbox +4\Fc_3\ofbox \right)\right]G^\m_\n
\\&& 
-4(\na^\m\nabla_\n-\d^\m_{\n}\Box)\bigg[\Fc_1\ofbox +\frac{1}{2}\Fc_2\ofbox +\Fc_3\ofbox \bigg]R = 0
\nn
\eea
The latter equation becomes identical to the one in GR, and therefore would give rise to only a two-polarization graviton, if and only if
\bea \label{IDG_recovers_GR_conditions}
&&\Fc_2\ofbox +4\Fc_3\ofbox \equiv 2 \Fc_\Weyl \ofbox =0
\\&&
\Fc_1\ofbox +\frac12\Fc_2\ofbox +\Fc_3\ofbox \equiv \Fc_R \ofbox + \frac13 \Fc_\Weyl \ofbox = 0
\nn
\eea
(note that having 1 instead of 0 in the latter equation would result in a Brans--Dicke scalar). 
The above conditions are equivalent to having only delocalized Gauss--Bonnet term \eqref{Gauss-Bonnet-generalized-nonlocal} in the action \eqref{action_IDG_Weyl_basis}.
Thus, this term can be set aside in the discussion of linearized perturbations.

We can replace the conditions \eqref{IDG_recovers_GR_conditions} with another set of conditions, which renders the linearized equations \eqref{IDG_EOM_linearized_Riemann_basis} different from those of GR,  but preserves the spectrum.
To this end, we choose the combination inside the brackets of the first term to be an exponent of an entire function
\be
\mP^{2} + 4\Box\Fc_C\ofbox = \mP^{2}e^{2\omega\ofboxstar } \, ,
\ee
resulting in
\be \label{FCnoghosts}
\Fc_C\ofbox = \mP^{2} \frac{e^{2\omega\ofboxstar }-1}{4 \Box} \, ,
\ee
while the absence of a scalar dictates
\be \label{proprel}
\Fc_R\ofbox +\frac{1}{3}\Fc_C\ofbox = 0 \, .
\ee
These conditions are equivalent to \eqref{ghost-free-no-sclarals-condition} and \eqref{ff_as_exp}.

Another possible choice is to introduce an additional scalar \cite{Koshelev:2022wqj}.
In this case, $\Fc_C$ remains as in (\ref{FCnoghosts}) but for $\Fc_R$ one gets
\be
\Fc_R \ofbox = \mP^{2} \frac{\left(  \Box/M^2 -1\right) e^{2\omega\ofboxstar}+1}{12 \Box} \, ,
\ee
where $M$ is the mass of the new scalar field.

\section{\label{sec:app:intermediate}Some details about the heat kernel technique}

\allowdisplaybreaks

The Q-functionals used in \eqref{universal_functional_trace_final} are defined as 
\best
\label{Q_functional_definition}
Q_m [f] := \int_0^\infty d s ~s^{-m} \tilde{f}(s) \, ,
\eest
where $\tilde{f}$ is the inverse Laplace transform of $f$.
Then, for $m$ positive integer we have
\best
\label{Q_functional_expr_pos}
Q_m [f] = \frac{1}{\Gamma(m)} \int_0^\infty dz ~z^{m-1} f(z),
\eest
whereas for negative integer we can choose $k$ such that $m+k>0$ and then
\best
\label{Q_functional_expr_neg}
Q_m [f] = \frac{(-1)^k}{\Gamma(m+k)} \int_0^\infty dz ~z^{m+k-1} f^{(k)}(z) \, , 
\eest
and $Q_0 [f] = f(0)$.
The first three invariants with even number of uncontracted derivatives that we used in  \eqref{universal_functional_trace_final} are expressed as
\bea \label{K_heat_kernel_invariants}
K^{(n)}(x) =&&\, \overline{a_n (\D)} \, ,\\\
K^{(n)}_{(\m\n)}(x) =&&\,
 -\frac{1}{2} g_{\m\n} \overline{a_n (\D)}
 + \overline{\na_{(\m}\na_{\n)} a_{n-1}} \, ,\nn\\
K^{(n)}_{(\m\n\r\l)}(x) = && \,
 \frac{3}{4} g_{(\m\n}g_{\r\l)} \overline{a_n (\D)}
- 3 g_{(\m\n} \overline{\na_\r \na_{\l)} a_{n-1} (\D)}
\nn\\&&
+ \overline{\na_{(\m} \na_\n \na_\r \na_{\l)} a_{n-2} (\D)} \, .
\nn
\eea
Here $a_n (\D)$ are the heat kernel coefficients
defined in a standard way, see, for example, \cite{Groh:2011dw}, and the overline stands for the coincidence limit.
The divergent contribution of \eqref{3rd_ghost_op} corresponding to the last term in \eqref{1-loop_EA_with_ghosts} is (recall that in our gauge $\g = 2/3 -1/x$, 
):
\best
\begin{aligned}
\label{EA_vector_Box_Q}
&\Tr\Log {\mathbb C} = \frac{1}{16 \pi^2} \log \left( \frac{\LUV^2}{\m^2} \right) \int\! d^4 x \;\sqrt{g}\, 
\times \\& \times 
\bigg[ 
\frac{(q+1) \left(\g^2 (10 q+37)-5 \g q-5 (q+1)\right) }{240 \g^2} \WeylSq  
\\ & 
+ \frac{(q+1) \left(\g^2 (q+7)+\g (4 q-6)-5 (q+1)\right)}{144 \g^2} R^2 
\\ & 
- \frac{(q+1) \left(\g^2 (30 q+67)-15 \g q-15 (q+1)\right)}{720 \g^2} \EulerInv
\bigg] \, .
\end{aligned}
\eest
The Faddeev--Popov ghost operator is
\best
\D_{\rm gh\,\m}{}^{\n}
= - \d_{\m}^{\n}\Box
- (1+2\b) \na_\m \na^\n
- R_\m{}^\n \, ,
\eest
which produces the contribution of
\best
\begin{aligned}
\label{EA_ghosts}
&\Tr\log \D_{\rm gh} = - \frac{1}{16 \pi^2} \log \left( \frac{\LUV^2}{\m^2} \right) \int\! d^4 x \;\sqrt{g}\, 
\times \\& \times
\Big[  
\frac{\left(8 x^2-150 x+45\right) }{540 x^2} \WeylSq 
+\frac{\left(80 x^2-96 x+45\right)}{324 x^2} R^2
\\ &
- \frac{\left(41 x^2-150 x+45\right)}{540 x^2} \EulerInv 
\Big] \, ,
\end{aligned}
\eest
corresponding to the second term in \eqref{1-loop_EA_with_ghosts}.

To perform the calculation, the \emph{xAct} collection of computer algebra packages, specifically \emph{xTensor} \cite{xAct:xTensor}, \emph{xPerm} \cite{Martin-Garcia:2008ysv}, \emph{xPert} \cite{Brizuela:2008ra}, 
\emph{Invar} \cite{Martin-Garcia:2007bqa, Martin-Garcia:2008yei}, \emph{SymManipulator} \cite{xAct:SymManipulator}, and \emph{xTras} \cite{Nutma:2013zea} were used.

\onecolumngrid
\section{\label{sec:app:integrals}Momentum Integration}

This section contains a proof that form factor logarithms, such as the Tomboulis one, as well as \eqref{omega_example_Alexey_1} and \eqref{omega_example_Alexey_2}, can be integrated using their leading UV asymptotics, when only UV divergences are of interest.
The subleading terms do not contribute to the beta functions.

Introducing the running scale $\m$ and the UV-cutoff as $\L$, we have the integrals as
\be \label{integrals_sum}
\Tr\log \Big[ \D^2 \cF (\D)) \Big]
+ \sum_{i=1}^\infty T_i \int_\m^{\LUV} d z \, I_i (z) \, ,
\ee
where
\be \label{integrals_list}
\begin{aligned}
& I_1 = 1/z 
\, , \quad
I_2 = 1/\cF(z)
\, , \quad
I_3 = \frac{1}{z \cF(z)} 
\, , \quad
I_4 = \frac{1}{z \left(\cF(z)\right)^2} 
\, , \quad
I_5 = - \frac{\cF'(z)}{\cF(z)}
\, , \\&
I_6 = - \frac{\cF'(z)}{\left(\cF(z)\right)^2}
\, , \quad
I_7 = - \frac{z \cF'(z)}{\cF(z)}
\, , \quad
I_8 = \frac{z \left(\cF'(z)\right)^2}{\left(\cF(z)\right)^2}
\, , \quad
I_9 =  \frac{z \cF''(z)}{\cF(z)}
\, .
\end{aligned}
\ee
Here, prime stands for the derivative over $z$,
the spectrum of the positive definite operator $\D = - \Box$ (in Euclidean signature).
The expressions $T_1, \dots, T_9$ are far too complicated to be expressed here, but can be found in shared files.
In what follows, we consider the high-energy limit 
$\LUV\to+\infty$.
We have 
\be \label{form_factor_leading_subleading}
\cF(z) = A z^q + \Fsub(z) \, ,
\ee
where $A\neq 0$, and $\Fsub$ is suppressed.
For the considerations below to be valid, it is sufficient to assume that
\be \label{subleading_condition}
\left\rvert \lim_{z\to+\infty} z \Fsub(z) \right\rvert < \infty \, ,
\quad\quad
\lim_{z\to+\infty} z^{1-\epsilon} \Fsub(z) = 0 \, ,
\ee
\ie\, the highest asymptotics of $\Fsub$ is $1/z$.
If that is the case, one can further separate the sub-subleading part as
\be \label{subleading_further_split}
\Fsub(z) = 1/z + \Fsubsub(z) \, .
\ee
For the form factors in the section \ref{subsec:general_ff}, $\cF_{\rm sub\,sub}$ is exponentially suppressed, and for the Tomboulis form factor, it is suppressed double-exponentially.
We want to prove that if only the UV divergences are of interest, substituting $Az^q$ is sufficient, and all other terms can be neglected.
Indeed, looking at the first term in \eqref{integrals_sum}, we have
\bear \label{trace_F_minus_trace_Box_q}
&\Tr\log \,\cF(\D) - \Tr\log \, \D^q 
= \Tr\log \Big[ \frac{\D^q + 1/\D + \Fsubsub(\D)}{\D^q} \Big]
\\&
= \Tr \Big[ \frac{1}{\D^{q+1}} + \frac{\Fsubsub(\D))}{\D^q} 
- \frac{1}{2} \left( \frac{1}{\D^{q+1}} + \frac{\Fsubsub(\D)}{\D^q} \right)^2 + \dots \Big] \, ,
\eear
where we expanded the logarithm for the high-energy modes of the d’Alembert operator.
Using \eqref{universal_functional_trace_final}, one can derive that, in particular
\be
\label{log_div_no_ders}
\Tr \left[\frac{1}{\D^k}\right]_\logdivsubscript = \frac{1}{16 \pi^2} \frac{1}{\Gamma(k)} \log \left( \frac{\LUV^2}{\m^2} \right) \int d^4 x \sqrt{g} \;\tr\; \overline{ a_{2-k}(\D)} \, ,
\ee 
for $k=1,2$, while for $k\geqslant 3$ the  l.h.s. is convergent.
Thus we conclude that for $q\geqslant 2$ \eqref{trace_F_minus_trace_Box_q} is convergent, and therefore,
\be
\Tr\log \Big[ \D^2 \cF (\D)) \Big] \bigg\rvert_\logdivsubscript 
= \Tr\log \Big[ \D^{q+2} \Big] \bigg\rvert_\logdivsubscript 
= - \frac{q+2}{2} \frac{1}{\left( 4 \pi \right)^2} \log \left( \frac{\LUV^2}{\m} \right) \int d^4 x \sqrt{g} \, \overline{a_2 \ofdelta} \, .
\ee
Now we proceed with the study of the integrals \eqref{integrals_list}.
The first trivially gives the logarithm.
We noticed that for the monomial form factor $z^q$ case, $I_1$, $I_5$, $I_8$, $I_9$ produce logarithmically divergent contributions, while others are convergent.
Consider the second integral,
\be
\int_\m^{\LUV} dz\, I_2 
= \int_\m^{\LUV} \frac{z\, dz}{A z^q + \Fsub(z)} 
= \frac{1}{A} \int_\m^{\LUV} \frac{dz}{z^{q+1}} \Big[ 1 - \Fsub(z) / z^{q+1} + \left(\Fsub(z) / z^{q+1} \right)^2 - \dots \Big] \,
\ee
where we have substituted \eqref{form_factor_leading_subleading}. 
Using the condition \eqref{subleading_condition}, we see that it gives no logarithmically divergent contribution for $q \geqslant 2$.
Therefore, we obtain that the integral $I_2$ gives a vanishing contribution.
Analogous considerations can be straightforwardly applied to $I_3$, $I_4$, and $I_6$.
Considering the next integral,
\be
\int_\m^{\LUV} dz\, I_{5}
= \ln\, \cF (z)\bigg\rvert_\m^{\LUV}
= \ln \frac{{\cal F}(\LUV)}{{\cal F}(\m)} \, .
\ee
Using \eqref{form_factor_leading_subleading}, we have
\be
\ln \cF(\LUV)
= \ln \left[ A \LUV^q \left( 1 + \frac{\cF_{{\rm sub}} (\LUV) }{A \LUV^q} \right) \right]
= \ln A + q \ln \LUV + \ln \left( 1 +\frac{\Fsub(\LUV)}{A \LUV^{q}} \right) \, ,
\ee
and therefore,
\be
\int_\m^{\LUV} dz\, I_{5}
= \ln \frac{\cF(\LUV)}{\cF(\m)} \bigg\rvert_{\logdivsubscript}
= q \ln \frac{\LUV}{\m} \, .
\ee
Furthermore,
\be
\int_\m^{\LUV} dz\, I_8 
= \int_\m^{\LUV} dz\, z \left(\frac{qAz^{q-1}+\Fsub'(z)}{Az^{q}+\Fsub(z)}\right)^{2}
= \int_\m^{\LUV} \frac{dz}{z} \left( \frac{q + {\Fsub'(z)}/{Az^{q-1}}}{1 + {\Fsub(z)}/{Az^{q}}} \right)^{2} \, .
\ee
Then, with \eqref{subleading_condition} we have
\be
\int_\m^{\LUV} dz\, I_8 \bigg\rvert_{\logdivsubscript}
= \int_\m^{\LUV} \frac{dz}{z}q^{2}=q^{2} \ln \frac{\LUV}{\m} \, .
\ee
Analogous considerations produce
\be
\int_\m^{\LUV} I_{9} \bigg\rvert_{\logdivsubscript} 
= q(q-1) \ln \frac{\LUV}{\m} \, .
\ee
The one that slightly stands out is
\be
\int_\m^{\LUV} dz\, I_7 
= - \int_\m^{\LUV} dz\, \frac{z \cF'(z)}{\cF(z)}
= - \int_\m^{\LUV} dz\, \frac{q + \Fsub'(z)/\left( Az^{q-1} \right)}{1 + \Fsub(z)/\left(Az^q \right)}
\stackrel{\cF(z) = z^q}{\to}
\LUV^2  \, .
\ee
This integral gives quadratic divergence for a monomial form factor, and it yields an additional logarithmic divergence if a subleading contribution of $z^{q-1}$ was present.
However, looking at \eqref{subleading_further_split}, we conclude that the integral $I_7$ does not produce any logarithmic divergence.

To summarize, for $q \geqslant 2$ only $I_1$, $I_5$, $I_8$, $I_9$ give contributions of the type $\log (\LUV)$ and only the leading asymptotics of the form factor contributes to it.

\twocolumngrid

\end{appendix}
\end{CJK*} 

\makeatletter  
\interlinepenalty=10000  
\bibliographystyle{jhep-mymod}
\bibliography{main}
\makeatother  

\end{document}